\documentclass[twocolumn,showpacs,preprintnumbers,amsmath,amssymb]{revtex4}

\usepackage{graphicx}
\usepackage{dcolumn}
\usepackage{bm}
\begin{document}  
\title{Spatially improved operators for excited hadrons
on the lattice}
\author{Tommy Burch$^a$, 
Christof Gattringer$^{a}$, 
Leonid Ya.\ Glozman$^b$, Reinhard Kleindl$^b$,
C.\ B.\ Lang$^b$, 
and Andreas Sch\"afer$^a$\\
(for the BGR [Bern-Graz-Regensburg] Collaboration)}
\vskip1mm
\affiliation{$^a$ Institut f\"ur  Theoretische Physik, Universit\"at Regensburg,
D-93040 Regensburg, Germany}
\affiliation{$^b$ Institut f\"ur Theoretische Physik, FB Theoretische Physik 
Universit\"at Graz,
A-8010 Graz, Austria}
\begin{abstract}  
We present a new approach for determining spatially optimized operators
that can be used for lattice spectroscopy of excited hadrons. Jacobi 
smeared quark sources with different widths are combined to 
construct hadron operators with different spatial wave functions. We
use the variational method to determine those linear combinations of
operators that have optimal overlap with ground and excited states. The 
details of the new approach are discussed and we demonstrate the power 
of the method using examples from quenched baryon and meson spectroscopy.
In particular we study the Roper state and $\rho(1450)$ and discuss 
some physical implications of our tests.
\end{abstract}
\pacs{PACS: 11.15.Ha}
\keywords{Lattice gauge theory, hadron spectroscopy}
\maketitle

\section{Introduction}\label{SectIntroduction}

Ground state spectroscopy on the lattice is by now a well understood
physical problem and impressive agreement with experiment has been achieved.
The lattice study of excited states is not so far advanced and for
many systems even the correct level ordering of the states has not 
yet been observed. A prominent example for such a system is the nucleon 
and its excited states. In particular the 
first excited positive parity nucleon, 
the so-called Roper state, has seen a lot of 
attention from the lattice community during the last two years
\cite{oldcalcs} - \cite{broemmeletal}.  Revealing the 
true nature of the Roper state with non-perturbative 
methods is an important task.

In a lattice calculation the masses of excited states show up in the 
subleading exponentials of Euclidean two point functions. A direct fit of 
a single Euclidean correlator is cumbersome since the signal is 
strongly dominated 
by the ground state. Also with methods such as constrained fits 
\cite{constrainedfits} or 
the maximum entropy method \cite{maximumentropy}
one still needs very high statistics for 
reliable results.

An alternative method is the computation of not only the correlator of
a single operator, but the calculation of a full matrix containing all cross
correlations of a set of several operators with the correct quantum numbers
\cite{variation}.
In this so-called variational method the correlation matrix is then 
diagonalized and one can show that each eigenmode is dominated by a different
physical state (for a properly chosen set of basis operators). 
After normalization at distance 0, for $t > 0$ 
the largest eigenvalue gives the Euclidean correlator of 
the ground state, the second-largest eigenvalue corresponds to the first 
excited state, et cetera.

The success of the variational method depends strongly on the choice 
of the basis operators. They need to be linearly independent and
should give rise to a large overlap with the physical states. 
A physical hadron state has several characteristics, 
in particular a specific Dirac structure 
and its spatial wave function. Therefore it is advantageous to optimize the 
spatial properties of the interpolating operators. An example for this fact 
is the mentioned Roper state where the variational method based
on nucleon operators that differ only in their diquark content but have
the same spatial wave function did not lead to success \cite{broemmeletal}.
It can be argued that a node in the radial wave function is necessary to 
capture reliably the Roper state or other radially excited hadrons.

In lattice calculations it is important to devise an efficient way of
implementing a spatial wave function at low numerical cost. When the
quark wave function extends over several lattice sites, a naive approach 
would require the inversion of the Dirac operator on point sources located 
at all possible quark positions (see \cite{wavefunction}
for a discussion of such wave functions). 
Creating a wave function for quarks in this 
way quickly becomes numerically expensive. For improving the ground 
state wave function an effective technique, so-called Jacobi 
smearing, has been shown to give good results \cite{jacobi}. 
Here, a point-like quark source is smeared to a shape similar to
a Gaussian, i.e.\ the origin of the source is connected to neighboring 
lattice sites within a time slice by gauge transporters. 
Such a source increases the overlap
of the physical hadron states with the lattice operators used to
create these states and 
considerably reduces the fluctuations of hadron propagators.

In this article we demonstrate that combining Jacobi smeared quark sources
with {\it different} widths in the variational method provides a powerful
tool for the analysis of excited hadron states. After presenting the outlined
ideas in detail we apply our method to quenched baryon and meson 
spectroscopy.  The excited nucleon system as 
well as excited mesons are analyzed (for recent lattice studies of
excited mesons see \cite{mesons}). We find good effective 
mass plateaus for the first and partly the second radially
excited states. The propagators can then be fitted 
using standard techniques. 
Different physical ramifications and implications of our findings are briefly
discussed.  

\section{The method}\label{SectMethod}

Let us begin the presentation of our method with a brief recapitulation of
Jacobi smearing of quark sources. Since a complete quark propagator (the
inverse of the lattice Dirac operator $D$) is a far too large object to be 
stored completely in the computer, one has to work with the propagator
evaluated on some source $s^{(\alpha, c)}$. The source is placed at timeslice 
$t = 0$ of the lattice. It is labeled by
a Dirac index $\alpha$ and a color index $a$. One then computes
($\rho$ and $c\,$ are summed over)
\begin{equation}
r^{(\alpha, a)} \, (\vec{x},t)_{\beta \atop b} \; \; = \; \; \sum_{\vec{y}}
D^{-1}(\vec{x}, t \mid \vec{y}, 0)_{\beta \, \rho \atop b \, c} \; 
s^{(\alpha, a)} (\vec{y},0)_{\rho \atop c} \; .
\label{diraconsource}
\end{equation}
When one chooses a point-like source at the spatial origin $\vec{0}$, i.e.\ 
$s = s_0$ with 
\begin{equation}
s^{(\alpha, a)}_0 (\vec{y},0)_{\rho \atop c} \; \;  = \; \;
\delta(\vec{y} - \vec{0}) \; \delta_{\rho \, \alpha} \; \delta_{c \, a} \; ,
\label{pointsource}
\end{equation}
the resulting vector is the quark propagator from the origin to all
lattice points. These quark propagators can then be combined 
to form hadron propagators. 

Choosing a point-like quark source has, however, the big disadvantage 
of a poor overlap with the true wave function. 
After all, for the lattice spacings which are appropriate,  
quarks are not expected to be located at a 
single point inside the hadron and the overlap of the point-like
wave function with the true physical wave function is small. The situation 
can be improved, e.g.\ by Jacobi smearing. One acts with a smearing operator 
$M$ on the point-like source $s_0^{(\alpha,a)}$ to obtain the smeared source 
$s^{(\alpha,a)}$:
\begin{eqnarray}
s^{(\alpha,a)} & \; = \; & M \; s_0^{(\alpha,a)} 
\; \; \; , \; \; \; \; 
M \; = \; \sum_{n=0}^N \, \kappa^n \, H^n \; ,
\nonumber
\\
H(\vec{x},\vec{y}\,) & \; = \; & \sum_{j = 1}^3
\Big[ \, U_j(\vec{x},0) \, \delta(\vec{x} + \hat{j\,}, \vec{y}\,) 
\nonumber
\\
&& \; \; \;  + \; \,
 U_j(\vec{x} \! - \! \hat{j\,},0)^\dagger \, 
\delta(\vec{x} \! - \! \hat{j\,}, \vec{y}\,) 
\, \Big] \; .
\label{jacobismear}
\end{eqnarray} 
The operator $H$ is simply the spatial hopping part of the 
Wilson term at timeslice 0. Note that $H$, and thus $M$, 
is trivial in Dirac space and acts only on the color indices 
(in our notation we suppress the color indices of $M$, $H$ and of the 
gauge transporters $U$). 

\begin{figure}[t!]
\vspace*{1mm}
\hspace*{-3mm}
\includegraphics*[height=5.4cm]{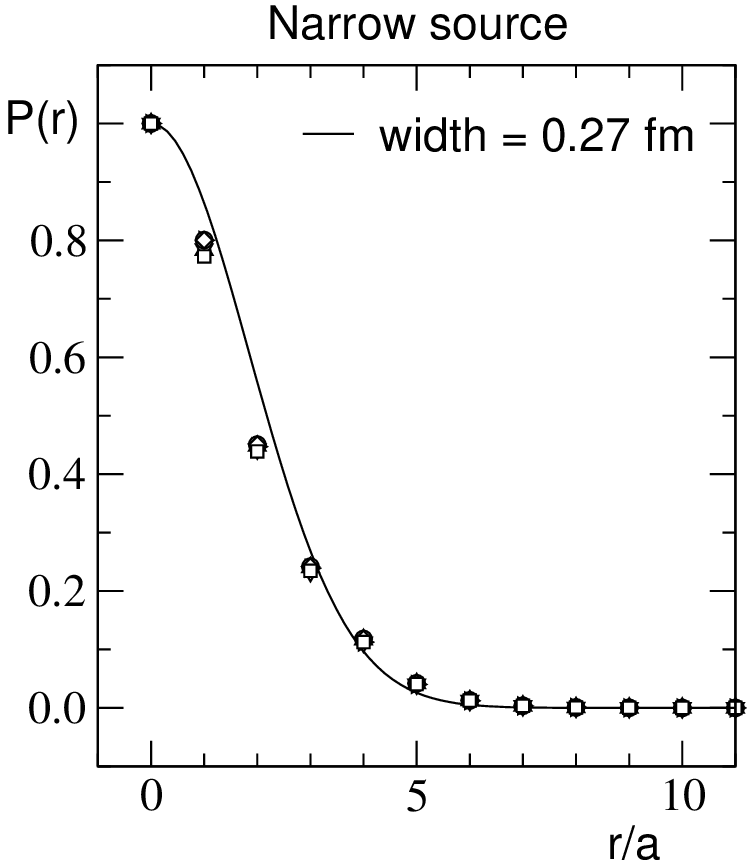} 
\hspace{-2mm}
\includegraphics*[height=5.4cm]{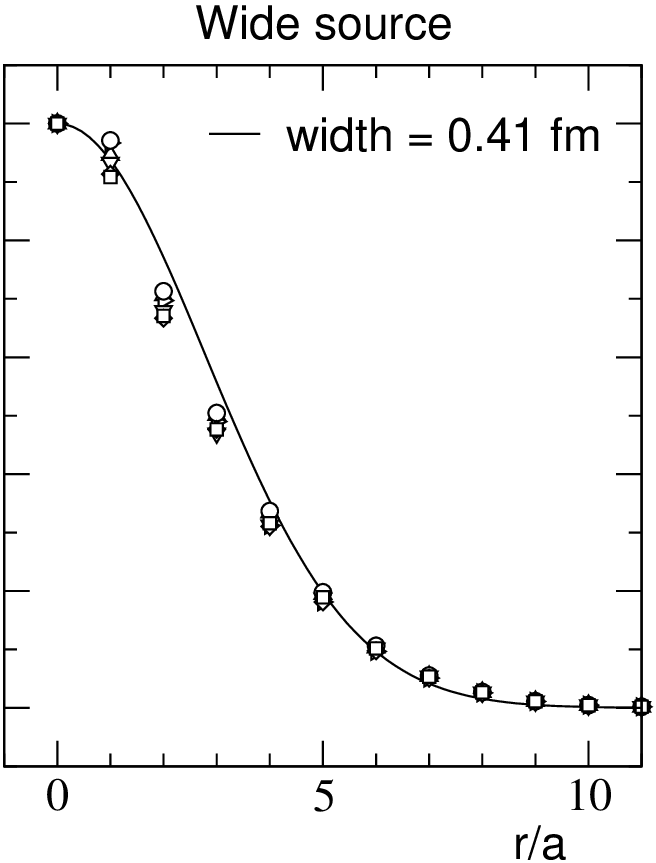}
\caption{Profiles $P(r)$ of the narrow and wide source. 
The symbols are our data points, the curves
are the target Gaussian distributions which we approximate by 
the profiles $P(r)$.
\label{fig1}}
\end{figure}

The Jacobi smearing outlined in Eq.\ (\ref{jacobismear}) has two 
free parameters: The number of smearing steps $N$ and the positive
real parameter $\kappa$. These two parameters can be used to adjust the
profile of the source. In order to study the profile of the smeared 
source $s^{(\alpha,a)}$ we define 
\begin{equation}
P(r) \; = \; \sum_{\vec{y}} \delta \Big( \, |\vec{y} \,| -  r\, \Big) \, 
\sum_b \, 
\Big| s^{(\alpha, a)} (\vec{y},0)_{\alpha \atop b} \Big|\; .
\label{profile}
\end{equation}
Since the smearing is trivial in Dirac space the profile 
function $P(r)$ is independent of the Dirac label $\alpha$ which appears
in the source $s^{(\alpha, a)}$. The profile function $P(r)$ can, however, 
depend on the color label $a$, but as we will demonstrate below this
dependence is small. The delta function on the right hand side is 
implemented by binning the discrete values of $|\vec{y\,}|$ on the
lattice. We stress that $P(r)$ is not a gauge invariant object,
but only an auxiliary quantity to visualize the source. Certainly it is 
easy to construct a profile function which is a color singlet, but since
we are interested in visualizing the sources $s^{(\alpha,a)}$ individually
for all $a = 1,2,3$, we use the form Eq.\ (\ref{profile}).

In this article we work with two different sources, a narrow source $n$
and a wide source $w$ with parameters given by 
\begin{eqnarray}
\mbox{narrow source} \; \; n \; : \; \; N  = 18 \, , \, \kappa = 0.210
\nonumber
\\
\mbox{wide source} \; \; w \; : \; \; N = 41 \, , \, \kappa = 0.191
\label{smearparams}
\end{eqnarray}
In Fig.\ \ref{fig1} we show the profile functions 
$P(r)$ for these two sources. 
They were normalized such that $P(0) = 1$. For
two different gauge configurations ($20^3\times 32$, L\"uscher-Weisz gauge
action, $\beta = 7.90$, lattice spacing $a = 0.148$ fm) 
we superimpose the values for $P(r)$ for 
all three possible color indices of the original point 
source, i.e.\ for $a = 1,2,3$.
It is obvious from the plots that the different color components of the 
point source lead to very similar profiles for the smeared source. Also
when comparing different configurations, the relative fluctuations are small. 
The smearing parameters $N$ and $\kappa$ were chosen such that the profiles
approximate Gaussian distributions with widths of $\sigma \sim 0.27$ fm 
for the 
narrow source and $\sigma \sim 0.41$ fm for the wide source. These Gaussians
are displayed as curves in Fig.\ \ref{fig1}. 

We remark that the parameters were chosen such that simple linear 
combinations $c_n \, n \, + \, c_w \, w$ 
of the narrow and wide profile approximate the
first and second radial wave functions of the spherical harmonic
oscillator: The coefficients $c_n \sim 0.6, c_w \sim 0.4$ approximate a 
Gaussian with a width of $\sigma \sim 0.33$ fm, while the combination 
$c_n \sim 2.2, c_w \sim -1.2$ approximates the corresponding excited radial 
wave function with one node. 
Thus, the two sources we include allow the system to build 
up radial wave functions with and without a node.

The final form of the wave function is, however, not put in by hand, but 
is determined by the system through the variational method
\cite{variation}. In this
approach one does not calculate a single correlator, but instead a 
complete correlation matrix of operators $O_i, i = 1,2, \, ... \, R$ 
that create from the vacuum the state which one wants to analyze.
One calculates all cross correlations
\begin{equation}
C(t)_{ij} \; = \; \langle \, O_i(t) \, O_j^\dagger(0) \, \rangle \; .
\label{corrmatdef}
\end{equation}
In Hilbert space this correlation matrix has the representation (for infinite 
temporal extent)
\begin{equation}
C(t)_{ij} \; = \; \sum_n \langle \, 0 \, | \, O_i \, | \, n \, \rangle 
\langle \, n \, | \, O_j^\dagger \, | \, 0 \, \rangle \, e^{-t \, M_n}  \; ,
\label{corrmatrix}
\end{equation}
where the sum runs over all physical states $| n \rangle$ and the 
corresponding energies are denoted as $M_n$. The eigenvalues 
$\lambda^{(k)}(t)$ of the correlation matrix can be shown to behave as
\begin{equation}
\lambda^{(k)}(t) \; \propto \; e^{-t \, M_k} \,[ \, 1 +
{\cal O}(e^{-t \, \Delta M_k}) \,] \; ,
\label{eigenvaluedecay}
\end{equation}
where $\Delta M_k$ is the distance of $M_k$ to nearby energy levels.
A modification of the method is the analysis of the generalized
eigenvalue problem
\begin{equation}
C(t) \, \vec{v} \; \; = \; \; \lambda(t) \, C(t_0) \, \vec{v} \; ,
\label{generalized}
\end{equation}
which can be rewritten to a standard eigenvalue problem by bringing
$C(t_0)$ to the left hand side of Eq.\ (\ref{generalized}), e.g.\
in the form of $C(t_0)^{-1}$ multiplied from the left. The normalization
at some slice $t_0 < t$ is expected to improve the signal by 
suppressing the contributions of higher excited states. We will explore
the freedom of such a normalization in our analysis.

The sources we use for the correlation matrix are constructed from the 
narrow and wide quark sources we prepared. This is best explained in an 
example: An operator which creates the $\rho$-meson is 
given by $\overline{u} \gamma_i d$. Both the $u$ and the $d$ quark can 
either have a narrow ($n$) or a wide ($w$) quark source. This gives the
four possible combinations $(n,n), (n,w), (w,n), (w,w)$, where the 
first entry refers to the smearing of the $u$ quark and the second 
entry is for the $d$ quark. Thus, we can use a basis of the four operators
\begin{equation}
{\cal O}_1 = (n,n) , \; {\cal O}_2 = (n,w) , \; 
{\cal O}_3 = (w,n) , \; {\cal O}_4 = (w,w) \; ,
\label{sm}
\end{equation}
for building up the correlation matrix $C(t)$. We remark that the 
different operators ${\cal O}_j$ also have to be used at the sink end
(timeslice $t$). This can be implemented by applying the smearing operator
$M$ with the two sets of smearing parameters Eq.\ (\ref{smearparams}) at 
the sink end of the quark propagator. 

Before we put our method to a test in the nucleon and meson systems
let us briefly summarize some technical details. 
For our quenched calculation we use the chirally
improved Dirac operator \cite{chirimp}. It is an approximation of a
solution of the Ginsparg Wilson equation \cite{GiWi82}
which governs chiral symmetry on the lattice. The chirally improved Dirac
operator is well tested in quenched ground state spectroscopy \cite{bgr} 
where pion masses down to 250 MeV can be reached at a considerably smaller 
numerical cost than needed for exact Ginsparg Wilson fermions. For ground
states the chirally improved action shows very good scaling behavior. 

The gauge configurations we use for testing our method were generated on a 
$12^3 \times 24$ lattice with the L\"uscher-Weisz action \cite{Luweact}.
The inverse gauge coupling is $\beta = 7.9$, giving rise to a lattice
spacing of $a = 0.148(2)$ fm as determined from the Sommer parameter in
\cite{scale}. The statistics of our ensemble is 100 configurations. We
use 10 different quark masses $m$ ranging from $am = 0.02$ to 
$am = 0.20$.

\section{Example A: Excited nucleons}\label{SectNucleons}

The first example where we put our new method to a test is
the spectroscopy of excited nucleons. 
Spectroscopy of the lowest positive and negative
parity states is a key to understanding the physics of the
nucleon. From a lattice perspective the nucleon system still holds
a few unresolved puzzles and new methods will help to obtain 
a more complete picture.

It has long been noted that the observed ordering
of the lowest positive, $1/2^+, N(1440)$, and negative
parity excitations of the nucleon, $1/2^-, N(1535)$ is 'unnatural'. 
Indeed, a physical picture based on
linear confinement, Coulomb and color-magnetic terms,
always arranges the first radial excitation above the first
orbital excitation, i.e.\ the excited states have alternating parities. 
This outcome is in contrast to the 
observed nucleon masses, where the first radial excitation, $N(1440)$, is well
below the first orbital one, $N(1535)$. This has
prompted speculations that perhaps the Roper
resonance is not a three quark state, but a collective
excitation of the bag surface \cite{BROWN}, a gluonic state 
\cite{CARBURKISS}, a $N\sigma$ coupled
channel effect \cite{KREWALD}, a resonance in the pion -
skyrmion system \cite{KARCOH}, and most recently -
a pentaquark state with a scalar diquark - scalar diquark -
antiquark structure \cite{JW}. 

On the other hand, we have to expect that close to the chiral limit 
effects of the spontaneous breaking of chiral symmetry 
should be important. It has been suggested in \cite{GR} 
that in the low-lying baryons the residual interaction between the
valence constituent quarks mediated by the Goldstone boson
field should be of vital importance. Such an interaction
is of the flavor- and spin-exchange nature, contrary to the
perturbative QCD degrees of freedom, and very naturally
resolves the puzzle of low-lying baryon spectroscopy
in the u,d,s sector. 

If chiral symmetry breaking is important for the Roper state,
we expect that the ordering of the lowest 
excitations of baryons should change as a function
of the quark mass. Baryons made of heavy quarks, where chiral effects 
do not play a role, should show a radial excitation above the
orbital excitation, while for smaller quark masses we expect
a level reordering as seen in the nucleon system. At intermediate
quark masses a level crossing of the radial and orbital excitations 
should take place. 
Hence, studying the evolution of the baryon spectrum versus
the current quark mass allows one to clarify the physical
picture. This can be done within the lattice
approach where  masses of quarks are external
parameters which can be freely varied. In particular with the
newly developed fermion actions based on the Ginsparg-Wilson equation
the region of small quark masses also becomes accessible.

Previous attempts to study these issues on the lattice
\cite{oldcalcs} - \cite{broemmeletal} have mainly used the two interpolating
fields, $\chi_1$ and $\chi_2$, 
\begin{eqnarray}
\chi_1(x) & \; = \; &
\epsilon_{abc} \left[u_a^T(x) \, C \gamma_5 \, d_b(x)\right ] u_c(x) \; ,
\label{chi1}
\\
\chi_2(x) & \; = \; &
\epsilon_{abc}\left[u_a^T(x) \, C \, d_b(x)\right ] \gamma_5 u_c(x) \; .
\label{chi2}
\end{eqnarray}
Here $C$ is the charge conjugation matrix and $a,b,c$
are the color indices. For the Dirac indices we use matrix 
notation. While the coupling of
the operator $\chi_1$ to both the nucleon and the
lowest negative parity state has been reliably established,
the operator $\chi_2$ couples neither to the nucleon,
nor to the Roper state. For a detailed study of this
issue and for an interpretation of this fact see \cite{broemmeletal}.
Hence a remaining possibility to see the Roper state is
to use $\chi_1$ and try to separate the strong signal
of the ground state (the nucleon) from the weak signal of the
radial excited state (the Roper state, if indeed the
Roper state is a 3-quark state) at small Euclidean time.
This strategy has been followed in Refs.\ \cite{dongetal} and 
\cite{sasaki} where in the former case a multi-exponential
fit of the diagonal $\langle \bar \chi_1 \chi_1 \rangle$
correlator has been performed (using constrained curve
fitting \cite{constrainedfits}), while in the latter
the maximum entropy method \cite{maximumentropy} has been used. Both
papers report the observation of the Roper resonance and level
crossing towards the chiral limit. However, the two papers 
contradict each other on the value of the pion mass where this
level switching takes place (300-400 MeV versus 600 Mev).
Further improvement of the lattice technology is necessary to deepen
the understanding of the nucleon system from the lattice.

Both the open physical and technical questions of the nucleon system 
make it an ideal testing ground for our new approach. The goal is to
see whether this method allows one to identify a reliable signal for the
excited positive parity nucleon. Our analysis is based on the interpolator
$\chi_1$ defined in (\ref{chi1}). It contains
three quarks and each of these quarks can be smeared either narrow
($n$) or wide ($w$). This gives 8 possible combinations 
\begin{equation}
{\cal O}_1 = (n,n,n), \; {\cal O}_2 = (n,n,w), \; ... \; {\cal O}_8 = (w,w,w)
\; .
\end{equation}
In this notation the first entry refers to the left-most quark in Eq.\
(\ref{chi1}), i.e.\ the u quark inside the diquark part of $\chi_1$.
We remark that 'diquark' does not refer to a true clustering of quarks in the 
nucleon but is used for the combination of the first two quark operators 
in the interpolators $\chi_1, \chi_2$.
The second entry is for the $d$ quark and the third entry for the other 
$u$ quark outside the diquark in $\chi_1$. From these operators we calculate
the correlation matrix
\begin{equation}
C_{ij}^\pm(t) \; = \; \langle \, \overline{{\cal O}}_i(t) \, \frac{1}{2}
[1 \pm \gamma_4] \,
{\cal O}_j(0) \, \rangle \; , 
\end{equation}
where we have inserted projectors to positive and
negative parities. Using the relation $C^+(t) = - C^-(T-t)$, where
$T$ is the total time extent of our lattice, we combine the two correlators 
to improve the statistics. This gives rise to the combined correlator
$C(t)$ which we then use in the variational method. The signal for 
positive parity states is obtained for small $t$ when running forward 
in time, while the negative parity states propagate backward in time
($T-t$). Near $T/2$ there is a crossing region where the propagators of the
two parities mix. We focus on the positive parity states and
show plots of propagators and effective masses only up to about $T/2$.

The correlation matrices $C(t)$ are real and symmetric within error
bars and we symmetrize the matrices by replacing $C_{ij}(t)$ with
$[ C_{ij}(t) + C_{ji}(t) ]/2$. Subsequently, we calculate the eigenvalues
$\lambda^{(k)}(t)$ for all $t$. At each time\-slice $t$ we order 
the (real) eigenvalues, such that $\lambda^{(1)}(t)$ is the 
largest eigenvalue, $\lambda^{(2)}(t)$ is the second largest eigenvalue
et cetera. We remark that it has been suggested \cite{variation}, that
one should analyze the eigenvalues of the generalized 
eigenvalue problem Eq.\ (\ref{generalized}), e.g.\ by studying the
eigenvalues of the normalized correlation matrix
$C(t_0)^{-1} C(t)$ for some $t_0 < t$. This normalization is expected 
to reduce the uncertainties due to the admixture of higher excitations.
We do not confirm
such an improvement of the signal. We will discuss this normalization
below and for now continue the discussion using eigenvalues from the
unnormalized matrix $C(t)$.

An important issue is the choice of the operators that are 
included in the analysis. Too few operators may not be sufficient to span 
all physical states we want to analyze. Too many operators drive up the
numerical cost without necessarily improving the signal. If a newly added
operator creates a state which only has small overlap with true physical
states then it will not contribute to improving the signal. On the contrary, 
it will contribute noise to the correlation matrix and even decrease the
quality of the signal. When exploring our new method we analyzed correlation
matrices starting from size $2 \times 2$ all the way up to the maximal size
of $8 \times 8$. We find that the results for the masses of 
the ground state and the first excited state
agree within one standard deviation (s.d.) 
when comparing different combinations of
operators. We observe that when using more than 4 basis operators the signal 
for the lowest two states does not improve any further. 

For correlation 
matrices of four operators we systematically analyzed the possible 
combinations. For some sets of operators, such as e.g.\ 
$(n,n,n)$, $(n,n,w)$, $(n,w,n)$, $(w,n,n)$ we found that the effective 
masses from the first and second excited states are degenerate within
error bars. Indeed, the operators $(n,w,n)$ and $(w,n,n)$ must equally
couple to the same physical state. This is because the two quarks
within the brackets of (\ref{chi1}) form a scalar-isoscalar diquark 
and can always be interchanged. Hence an unnecessary repetition
of equivalent operators, like $(n,w,n)$ and $(w,n,n)$ should be avoided.
When working with the set
\begin{equation}
(n,n,n), \; (n,n,w), \; (n,w,n), \; (w,w,n) \; ,
\label{optset}
\end{equation}
which does not contain equivalent operators,
we find, at least for large quark masses, a splitting of the effective masses
from the second and third eigenvalues larger than one s.d. 

To summarize the
comparison of different sets of operators we find that the masses
of the ground and excited states  
agree within error bars. A distinction of the masses from the second and third 
eigenvalues is possible only for particular sets of operators which typically 
contain also operators with two wide sources 
(e.g.\ the set in Eq.\ (\ref{optset})).

\begin{figure}[t]
\vspace*{2mm}
\hspace*{-4mm}
\includegraphics*[width=7.5cm]{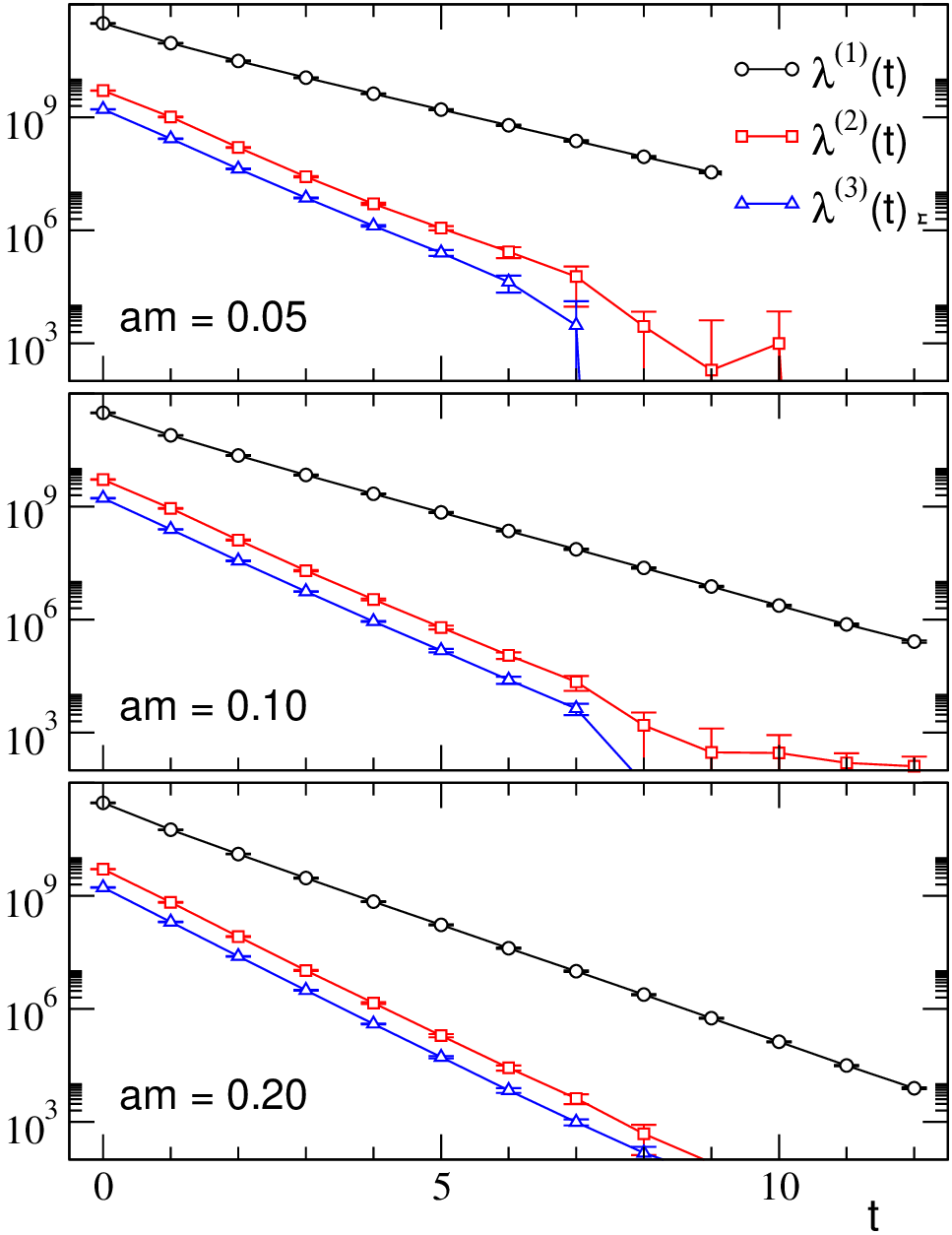}
\caption{ The three largest eigenvalues $\lambda^{(1)}$ (circles), 
$\lambda^{(2)}$
(squares) and $\lambda^{(3)}$ (triangles) for the quark masses $am=0.05$
(top), $am=0.10$ (middle) and $am=0.20$ (bottom).
\label{fig2}}
\end{figure}

In Fig.\ \ref{fig2} we show the positive parity parts of the three largest
eigenvalues $\lambda^{(1)}(t)$, $\lambda^{(2)}(t)$ and 
$\lambda^{(3)}(t)$ from the
$4 \times 4$ correlation matrix with the set of operators 
(\ref{optset}) constructed from  
the local $\chi_1$ interpolator. The exponential
decay of all three eigenvalues is clearly seen and 
the slopes differ, most obviously for the ground state and the 
first excited state. We identify these
signals with the nucleon, the Roper state and 
the next positive parity resonance $N(1710)$.
We remark that in the heavy quark region
the latter two states belong to the same shell and hence must be
approximately degenerate. However, given the fact that the
masses of the excited states extracted from the second and the
third eigenvalues are rather close, we perform an additional test 
that these eigenvalues represent different states.

It has been understood long ago that all Roper states form
an excited $\bf 56$-plet of SU(6). In this multiplet the
parity of any two-quark subsystem is positive. The $N(1710)$ belongs
to another multiplet, which contains both positive and negative
parity two-quark subsystems. The two-quark subsystem in the brackets
in $\chi_1$, Eq.\ (\ref{chi1}), has positive parity, 
while the subsystem in the brackets
of $\chi_2$ has negative parity. Hence, applying the same type
of smearings to the $\chi_2$ interpolator, we should see only the
signal from the $N(1710)$ state, and no signals from the nucleon and the
Roper. We have diagonalized a $4 \times 4$ correlation matrix with 
the source combinations listed in 
(\ref{optset}) applied to the local $\chi_2$ interpolator and 
indeed observed a signal only in one eigenvalue. This signal is 
clearly compatible with the $N(1710)$ signal obtained with $\chi_1$.

Before we discuss the effective masses and the fit procedures we 
used, let us address another important advantage of our source technique.
In \cite{dongetal} evidence for nucleon-$\eta^\prime$ ghost 
contributions (a quenching artifact) to the
nucleon correlators at small quark masses was given. These contributions
come with a negative coefficient and in  \cite{dongetal}
had to be included explicitly in the multi-exponential fitting function 
of the correlator. In our correlation matrix analysis
we find that the first and second eigenvalues are not at all affected 
by the ghost contributions, i.e.\ these correlators are positive and we 
observe undisturbed effective mass plateaus. Only the third and higher 
eigenvalues can become negative for large $t$. 
Thus, the ghost contributions are disentangled
from the ground state and the first excited state, i.e.\ the states 
we are interested in here.

\begin{figure*}[t]
\vspace*{1mm}
\includegraphics*[width=14.5cm]{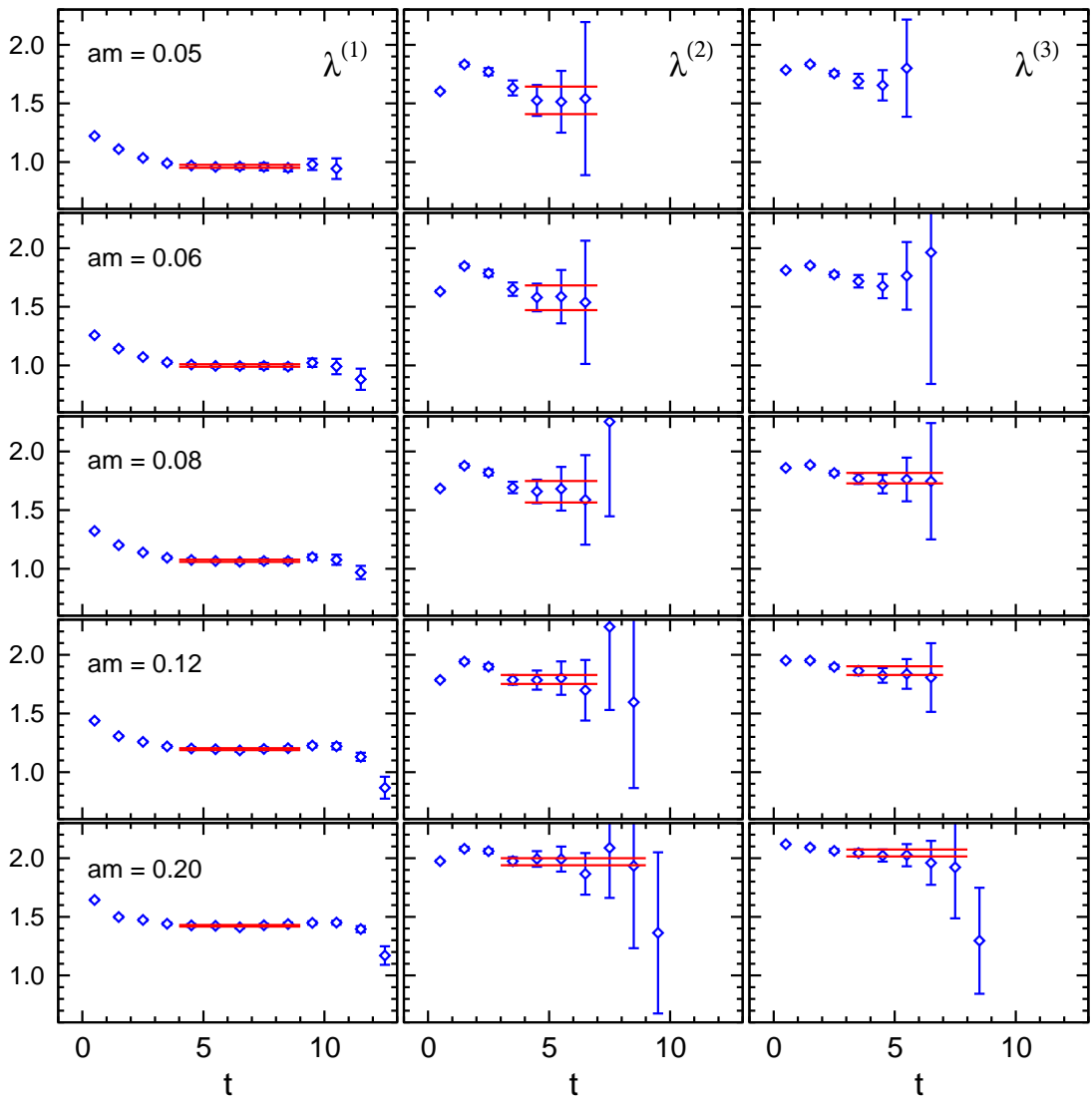} 
\caption{Effective masses from the first three eigenvalues (left to right)
for quark masses between $am = 0.05$ and $am = 0.20$ (top to bottom). The
symbols represent the numerical data. The horizontal lines
extend over the chosen fit range and represent the fit results plus and
minus the statistical error.  
\label{fig3}}
\end{figure*}

Let us now present the effective masses as obtained from the correlation 
of the operators listed in Eq.\ (\ref{optset}). In Fig.\ \ref{fig3} we 
show effective masses
\begin{equation}
m_{eff}^{(k)} \, \Big(t + \frac{1}{2}\Big) \; = \; 
\ln \left( \frac{\lambda^{(k)}(t)}
{\lambda^{(k)}(t+1)} \right) \; ,
\end{equation}
for the first three eigenvalues (out of a total of four) with
the data for $\lambda^{(1)}$ in the left hand side column, $\lambda^{(2)}$
in the center column and $\lambda^{(3)}$ on the right hand side.
We display the data for different quark masses from 
$am = 0.05$ up to $am = 0.20$ (top to bottom). The symbols show 
our numbers for the effective masses with statistical errors determined
with the jackknife method. The horizontal lines in the plots represent 
our fit results: We plot the one standard deviation (s.d.) error band for
the fitted mass. The band
extends over the $t$-interval which was used for the fit.
We stress that these are not fits to the effective mass, but fully correlated
fits to the propagators shown in Fig.\ \ref{fig2}. 
We discuss the details below.

The signal for the ground state appears in the effective mass for
$\lambda^{(1)}$ which we show in the left hand side column. We find 
well pronounced plateaus. The fit range was chosen from $t=4$ to 
$t=9$. For $t < 4$ we observe contributions from higher excited states,
while for $t > 9$ one enters the crossing region where contributions from 
the negative parity states, which propagate backwards in time, start 
to mix with the positive parity signal. 

The first excited state 
shows up in $\lambda^{(2)}$ (center column). It is obvious, that here the 
statistical errors are larger than for the ground state, in particular for 
$t > 4$. This, however, is as expected for a heavier state which has
a faster decreasing correlator, giving rise to a smaller signal to 
noise ratio already at not too large $t$. However, also for the first 
excited state we find good plateaus in the effective mass. These plateaus
extend from $t=3$ to $t=9$ for our largest mass $am = 0.20$ and shrink
to $t=4$ to $t=7$ for $am = 0.05$. We emphasize that there is a common
interval in $t$ where all 3 plateaus are seen simultaneously. 
It is very satisfactory to see a credible effective
mass plateau for the excited states
and not to have to rely on the fits alone to extract their masses. 

When inspecting the effective masses from the third eigenvalue one
finds again good effective mass plateaus. The statistical errors
are not larger than for the second eigenvalue and fits in similar $t$-ranges 
can be performed. We did not fit the $\lambda^{(3)}$ data for quark 
masses below $am = 0.08$ where we no longer can exclude contributions from
ghost states and observe a decreasing quality in the effective mass 
plateaus. When comparing the positions for the plateaus in the first
and second excited states we find that they are rather close to each other.
Only for the three largest masses, where the statistical errors are smaller,
can we establish a mass splitting larger than one s.d. Also in nature 
the mass splitting between the first and second excited positive parity 
states ($N(1440)$ and $N(1710)$) is relatively small. 

Before we come to presenting the final mass values of our analysis 
we need to discuss the procedures we used for fitting the correlators.
The exponential decay of the eigenvalues was fitted with the two parameter
ansatz $A \, \exp(\,-Mt\,)$.
Since the different values of $t$ are not statistically
independent we used fully correlated fits. The statistical errors and the 
covariance matrix were determined 
with the jackknife method.
The fit ranges were chosen such that they extend over
those values of $t$ where we see a credible effective mass plateau. Changing
the upper limit of the fit interval by $\pm 1$ does not affect the fit results 
(the variation is considerably less than one s.d.). Also an increase of the 
lower limit of the fit range changes the fit results by considerably less than 
one s.d. When decreasing the lower end of the fit interval, the result for the
mass typically goes up by one s.d. This effect is obvious from the effective 
mass plots where one sees that for smaller $t$ one runs
into the contributions of higher excited states. 
Our fits all have $\chi^2/d.o.f.$ values smaller than 1 
and the $\chi^2/d.o.f.$ 
also stays below 1 when considering the changes in 
the fit range. Thus, the $\chi^2/d.o.f.$ is not a very stringent criterion for
determining the fit range and our choice based on the effective mass 
plateaus is rather conservative. 

\begin{figure}[t]
\vspace*{1.5mm}
\hspace*{-5mm}
\includegraphics*[height=7.0cm]{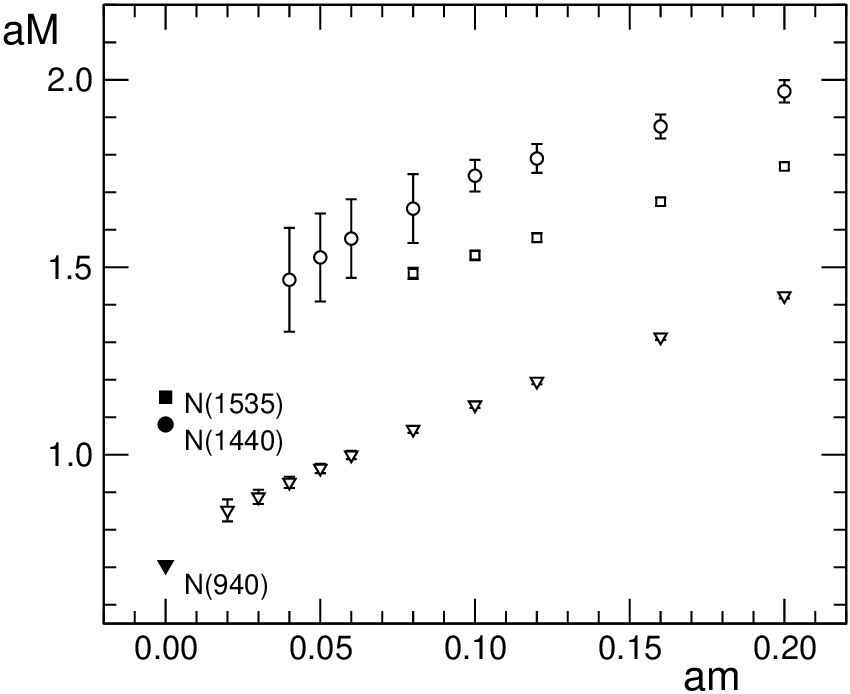}
\caption{Baryon masses as a function of the quark mass. The open symbols
represent our lattice results and the filled symbols show the experimental 
masses of the corresponding nucleons, converted into lattice units
with $a = 0.148$ fm as determined from the Sommer parameter. 
\label{fig4}}
\end{figure}

In our fits we carefully analyzed the possibility of normalizing our
matrix as suggested by the generalized eigenvalue problem 
Eq.\ (\ref{generalized}). In particular we tried the choices $t_0 = 0, 1, 2$
and found that the fits for the masses are unchanged within error bars. 
For the ground state the change was less than 0.3 s.d.\ and for the
excited state less than 1 s.d. For $t_0 = 2$ we 
find a noticeable decrease in the
quality of the effective mass plateaus, which comes from the fact the 
at $t_0 = 2$ the matrix $C(t_0)^{-1}$ which is used for the normalization 
already becomes affected by statistical noise. We cannot confirm any 
suppression of the effects from higher excited states and our 
results from the generalized eigenvalue problem are indistinguishable from 
the data obtained from the regular eigenvalue problem. In the further 
discussion we use the latter. 

We experimented also with another possibility to fit our correlators. Based 
on the Hilbert space decomposition (\ref{corrmatrix}) we used an ansatz of 
the form 
\begin{equation}
C_{ij}(t) \; = \; \sum_{n=1}^R  A^{(n)}_i \, A^{(n)}_j e^{-t\, M_n} \; ,
\label{multifit}
\end{equation}
to fit an $R \times R$ correlation matrix. The coefficients $A^{(n)}_i$ are 
chosen real, since the correlation matrix is real and symmetric. We find
that the ansatz (\ref{multifit}) can be applied successfully only for 
$2 \times 2$ matrices. For larger correlation matrices the data of our
ensembles used for testing the method are not accurate enough
for the multi-parameter fit (\ref{multifit}). The 
$\chi^2/d.o.f.$ becomes large and we sometimes encountered convergence 
problems in the numerical minimization of the $\chi^2$ functional. For the 
$2 \times 2$ case, however, the results from fitting the whole correlation 
matrix with (\ref{multifit}) agree reasonably well with the fits of the 
eigenvalues. For the ground state mass the two results differ by less than 
one s.d. For the first excited state the difference is below one s.d.\ 
for $am = 0.04$ rising to 1.5 s.d.\ at $am = 0.20$.
 
In Fig.\ \ref{fig4} we show the
fitted nucleon masses in lattice units as a function of the quark mass. Open 
triangles represent our data and the
statistical error for the ground state and the filled triangle gives
the experimental mass for the nucleon $N(940)$ (converted to lattice units
using the Sommer parameter). The open circles are the data
for the first excited state and the corresponding filled circle represents the 
mass of the Roper $N(1440)$. The open squares are the masses for the lowest
negative parity state calculated on the same ensemble of configurations 
in \cite{broemmeletal}. Note that the lower two data sets are for ground
states of positive and negative parity 
and thus have smaller statistical errors. Although we do not attempt 
a chiral extrapolation of our data (this is deferred to a future
large scale study of the nucleon system), our numbers seem to approach the
experimental data reasonably well. For the quark mass range in our 
study the first excited, positive parity masses still remain above the 
negative parity, ground state data, but a trend towards a level crossing 
is plausible. We remark that the smallest quark mass where we 
can fit the excited nucleon, $am = 0.04$, corresponds to a pion mass of 450 MeV 
(see \cite{bgr}). Thus our preliminary data contradict a level crossing 
at 600 MeV as claimed in \cite{sasaki}. Clearly further studies in 
larger volumes are important; such studies are in progress.

\section{Example B: Excited rho meson}\label{SectMesons}

As another test of our approach  we discuss the rho-meson $\rho (770)$
and its radial excitation $\rho (1450)$.
Traditionally the $\rho (770)$ has been studied with the
vector current interpolator,
\begin{equation}
\overline{u}(x) \, \gamma_i \, d(x) \; .
\label{vect}
\end{equation}
The same meson in the chiral symmetry broken regime
should also be seen with the time component of the tensor interpolator,
\begin{equation}
\overline{u}(x) \, \sigma_{4i} \, d(x) \; .
\label{tens}
\end{equation}
The difference between the two is that they transform under
different representations
with respect to $SU(2)_L \times SU(2)_R$ and $U(1)_A$ \cite{Gloz1}.
Again we use wide and narrow quark sources for both interpolators.
Thus, for both interpolators (\ref{vect}), (\ref{tens}) 
we can build the 4 operators as listed in Eq.\ (\ref{sm}). However,
the two combinations $(n,w)$ and $(w,n)$ give identical correlators
and one of them can be omitted. Thus, we evaluate two $3\times3$
correlation matrices, one for the vector interpolator and one for the
tensor interpolator (the different spatial components are averaged). 
For both these interpolators we check whether the 
ground and radially excited states couple. When we diagonalize
the $3 \times 3$ matrix with either the interpolator (\ref{vect}) or 
the interpolator (\ref{tens})
we see a pronounced exponential decay only for the two larger
(in magnitude) eigenvalues, $\lambda^{(1)}(t)$ and $\lambda^{(2)}(t)$.
The smallest eigenvalue $\lambda^{(3)}(t)$ 
does not show a clear effective mass plateau and
becomes negative at small quark masses for large $t$. This is
a clear indication that this eigenvalue couples to an unphysical
quenched ghost state \cite{Bardeen,DeGrand}. 

\begin{figure}[t]
\vspace{1mm}
\hspace*{-1mm}
\includegraphics*[width=8.6cm]{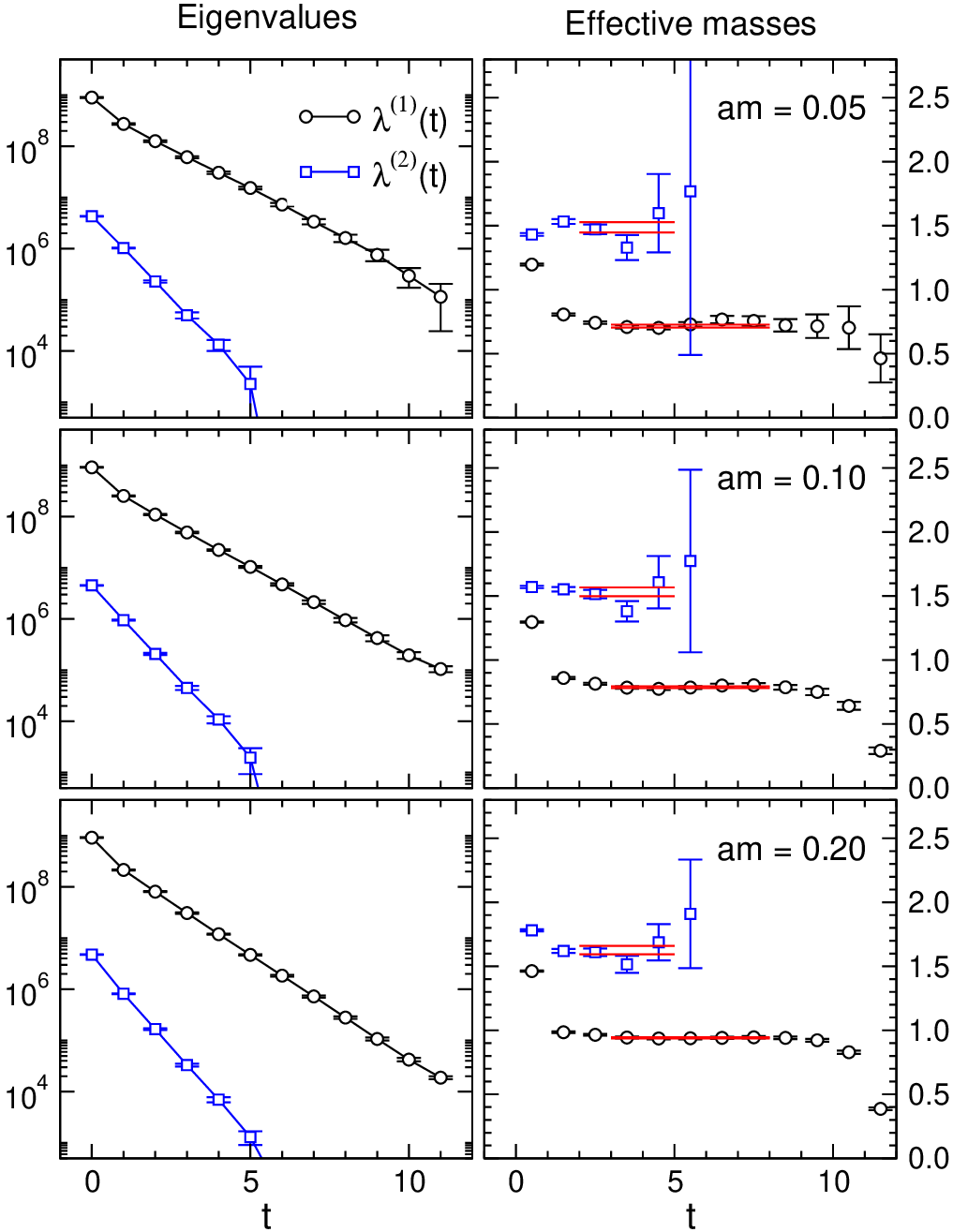}
\caption{Left hand side panels: The largest eigenvalue (circles) and the
next to the largest one (squares). Right hand side: The corresponding
effective mass plots. From top to bottom the quark masses are 
$am =0.05,0.10,0.20$.
\label{fig5}}
\end{figure}

In  Fig.\ \ref{fig5} we show the two larger eigenvalues 
and  the corresponding effective mass plots obtained with the 
tensor current (\ref{tens}) for three different quark masses
($am = 0.05, 0.10, 0.20$, from top to bottom). We observe 
a clean exponential decay for both eigenvalues and corresponding plateaus 
in the effective mass plots. We remark, that results for the mass obtained 
with the vector current (\ref{vect}) agree within the error bars. 
Contrary to the baryon case, we observe only one radial excitation.
Indeed, there is only a single one-node excitation for a meson
with the given quantum numbers.

\begin{figure}[b!]
\hspace*{-5mm}
\includegraphics*[width=8cm]{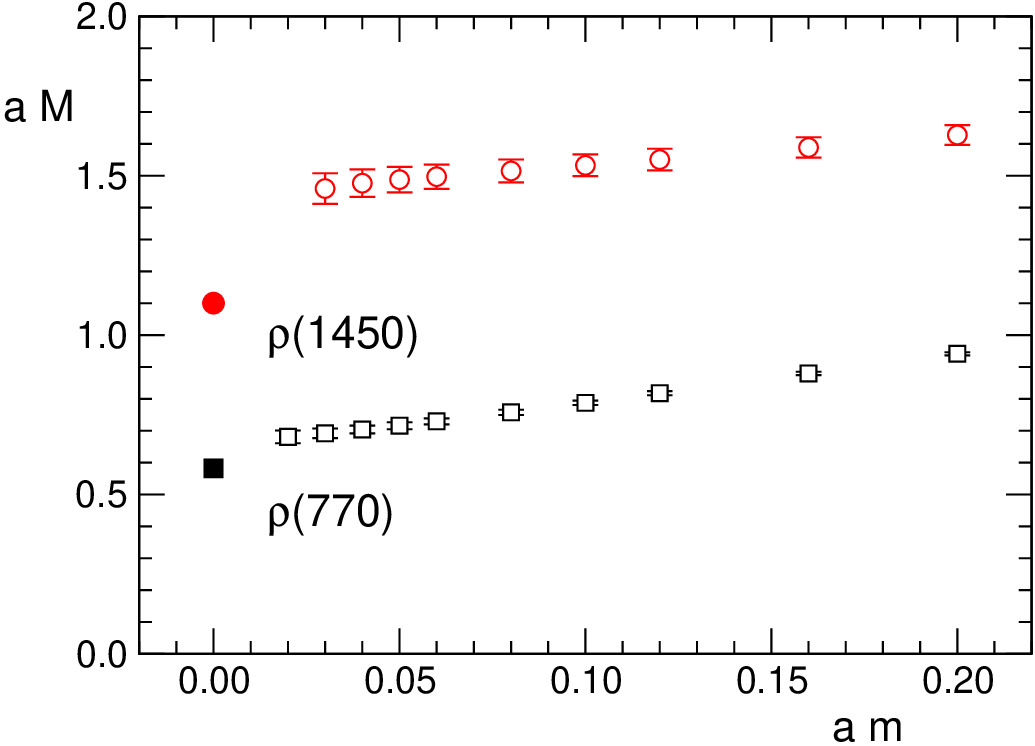}
\caption{Masses of $\rho(770)$ and $\rho(1450)$
as a function of the quark mass. 
\label{fig6}}
\vspace*{-2mm}
\end{figure}

Like for the baryons, we apply a standard correlated single 
exponential fit to extract the meson masses. For the ground
state the time interval is chosen to be (3,8), while for the
excited state we use (2,5). The final results for the masses as a function 
of the quark mass are shown in Fig.\ \ref{fig6}. We find that the ground state
approaches its experimental value reasonably well (the experimental data
were converted to lattice units with the Sommer parameter scale).

The excited state masses are
considerably above their experimental value. There is, however,  
a plausible reason for this behavior. The sizes of hadrons which
are not, or only weakly, affected by spontaneous chiral symmetry
breaking can be estimated from the known string tension, 
which is approximately 1 GeV/fm. Hence the size
of the $\rho$-meson is expected to be below
1 fm, while the size of $\rho(1450)$ should be approximately
1.5 fm. The size of our lattice is 1.8 fm,
which is clearly not enough for a precise measurement of
the $\rho(1450)$ mass. The finite size effect cannot 
be neglected for the excited state and shifts the measured
mass up as compared to the experimental value. 
A study of the rho system on larger lattices is in progress.

\section{Analyzing the operator content of the physical states}

We have based our new approach on the working hypothesis that,
when using the variational method, 
the excited states have a better overlap 
with a basis of operators that allow for a node in the spatial wave 
function. A crucial test of this assumption is to check whether 
indeed the ground state is built from a nodeless combination of our 
sources and the excited states do show nodes. 

This question can be addressed by analyzing the eigenvectors of the 
correlation matrix (\ref{corrmatdef}). Let us denote by $\vec{e}^{\,(k)}$
the $k$-th eigenvector of the correlation matrix, and its eigenvalue
is $\lambda^{(k)}$. Then we can define optimal operators 
$\widetilde{{\cal O}}_k$ by
\begin{equation}
\widetilde{{\cal O}}_k \; = \; \sum_j \, c_j^{(k)} \,{\cal O}_j \; ,
\end{equation}
with the mixing coefficients $c_j^{(k)}$ determined from the entries
of the eigenvector $\vec{e}^{\,(k)}$ via
\begin{equation}
c_j^{(k)} \; = \; \vec{e}^{\,(k)\, *}_j \; ,
\end{equation}
where the asterisk denotes complex conjugation. 
The correlation matrix of the optimal operators $\widetilde{{\cal O}}_k$
is diagonal, i.e.\ 
\begin{equation}
\langle \, \widetilde{{\cal O}}_k(t) \, \widetilde{{\cal O}}_l(0)^\dagger 
\, \rangle \; = \; \delta_{kl} \, \lambda^{(k)}(t) \; .
\end{equation}
Thus, the operator $\widetilde{{\cal O}}_1$ has optimal overlap with
the ground state, the operator $\widetilde{{\cal O}}_2$ has optimal
overlap with the first excited state etc. By analyzing the
coefficients $c_j^{(k)}$ we can thus learn about the structure of 
the state seen in the eigenvalue $\lambda^{(k)}$. 
In particular we can address the 
question whether there are nodes in the wave functions. As we have discussed 
in Section 2, a node occurs if there is a relative minus sign between 
the wide and the narrow quark source. For the nucleon system, where we
work with the set of operators listed in (\ref{optset}), our 
working hypothesis implies that all
coefficients have the same sign for the ground state, while for the
excited states we expect relative signs.

\begin{figure}[t]
\vspace*{1mm}
\hspace*{-4mm}
\includegraphics*[width=7cm]{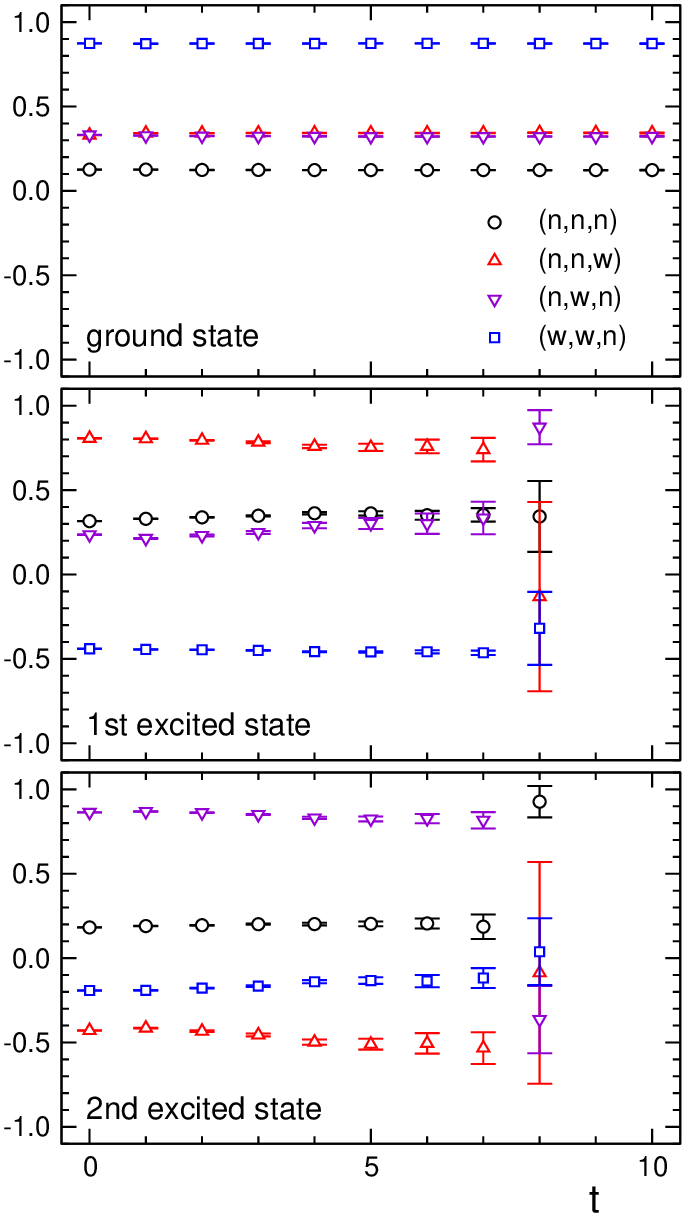}
\caption{Entries of the eigenvectors as a function of t. The data are for 
quark mass $am = 0.10$. The entries are labeled by the source combinations 
they correspond to. 
\label{fig7}}
\end{figure}

We determine the eigenvectors $\vec{e}^{\,(k)}$ of $C(t)$ at each 
value of $t$. Since $C(t)$ is real and
symmetric, the eigenvectors can be chosen real. In Fig.\ \ref{fig7} we 
display these eigenvectors for the nucleon analysis by plotting the 
(real) coefficients $c_j^{(k)} = \vec{e}_j^{\,(k)}$. In the top plot
we show the numbers for the ground state ($k = 1$), in the middle plot the 
numbers for the first excited state ($k = 2$) and in the bottom plot we
display the second excited state ($k = 3$). The different symbols 
represent the mixing coefficients for the different basis operators
and we use circles for $(n,n,n)$, triangles for $(n,n,w)$, 
upside down triangles for $(n,w,n)$ and squares for $(w,w,n)$, all 
referring to the interpolator $\chi_1$. The data we 
show are for quark mass $am = 0.10$. We remark that our eigenvectors
are normalized to one, implying that
\begin{equation}
c_1^{(k) \,2 } \, + \, c_2^{(k) \,2 } \, +  \, c_3^{(k) \,2 } 
\, + \, c_4^{(k) \,2 } \; = 1 \; ,
\end{equation} 
for the mixing coefficients. Furthermore, since $C(t)$ is real 
symmetric, the eigenvectors are also orthogonal to each other
for all $t$.

It is obvious from the plot that also the mixing coefficients
show plateaus. For the ground state these plateaus start at 
$t=0$, while for the excited states they typically start at $t = 4$, 
where we also observed the onset of plateau-like behavior in the effective 
masses. Starting at $t=8$ the plateaus for the coefficients vanish
for the excited states. At this $t$ the eigenvalues are already very 
small (they decay exponentially) and different eigenvectors start to mix.   
However, for all quark masses we observe long enough plateaus to 
clearly identify the mixing coefficients of the optimal operators. 

\begin{figure}[t]
\vspace{1.5mm}
\hspace*{-4mm}
\includegraphics*[width=7cm]{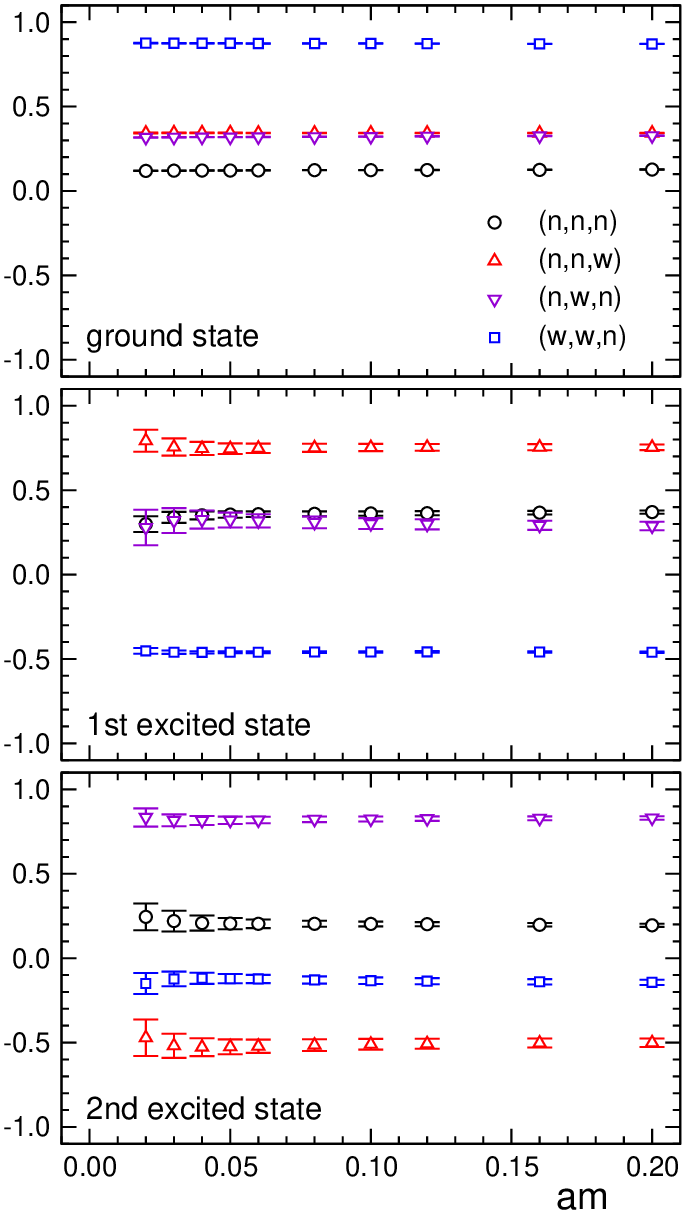}
\caption{Coefficients for the optimal operator combinations as a function of
the quark mass.
\label{fig8}}
\end{figure}

In Fig.\ \ref{fig8} we show the mixing coefficients as a function 
of the quark mass. In particular we choose the values of the coefficients 
for time slice $t = 5$. We use the same symbols for the different basis 
operators as in Fig.\ \ref{fig7} and again show from top to bottom 
the ground state, the first and the second excited state. 

The top plot shows 
that for the ground state (the nucleon) all 4 mixing coefficients 
have the same sign and the
wave function of the ground state does not have a node. For the two 
excited states the situation is clearly different. We find positive and 
negative coefficients, indicating that the wave function has a node. 
We stress that a linear combination of the profiles shown in Fig.\ \ref{fig1},
using the coefficients of Fig.\ \ref{fig8} should not be literally interpreted 
as the true nucleon wave function. Firstly, our narrow and wide sources are
not orthonormalized, and secondly they provide a much too small basis for 
mapping the details of a complicated three body wave function. 

We performed the same analysis also for the eigenvectors of the correlation
matrices we used for the rho meson. Again we confirm that the ground 
state has equal sign coefficients and thus is nodeless, while the excited state
has relative signs between its coefficients indicating a node. Thus
for both systems where we tested our method we could confirm that in 
the variational method combinations without node give rise to the 
ground state signal while the signal for excited states comes from 
combinations with a node.

\section{Summary}
We have presented a new approach to improving the spatial structure of 
operators used for hadron spectroscopy on the lattice. The central idea
is to combine Jacobi smeared quark sources of different width in the 
variational approach. This combination allows for nodes in the 
spatial wave function and improves the signal from radially excited states. 
We have demonstrated the power of this approach by
applying the method to a study of the lowest radial excitation of the nucleon,
the Roper resonance, and to the radial excitation of
the rho meson, $\rho(1450)$. In both cases we clearly
identify credible plateaus in the effective mass plots and
are able to trace the signal of these states from the heavy 
quark region towards the chiral limit. It is reassuring to see clear 
effective mass plateaus also for the excited states, and the corresponding
mass needs not be extracted from a multi-exponential fit. 
A simple single exponential fit is sufficient for extracting the mass from the
decay of the second eigenvalue. When plotting the fitted masses 
as a function of the quark mass, the agreement with the experimental situation 
is reasonable. A large scale study with larger lattices and a systematic 
chiral extrapolation is in progress.
We also note that it is possible to extend our method to the
next radial excitations, with two nodes. For that one needs to use
at least three different types of quark sources. We plan to study this 
extension of the method in the near future.

There are two physical implications of our preliminary observations:

(i) The fact that our numerical data for the first excited state can be 
plausibly interpreted as the Roper signal has an important consequence 
for the nature of this state: It implies that the Roper's leading Fock 
component is a 3 quark state. A trend towards a level switching of the 
lowest positive and negative parity excited states is  visible, which 
is consistent with the physical picture of Ref.\ \cite{GR}.
 
(ii) In our data the quark mass dependence of the excited rho meson state 
$\rho(1450)$ is only weak. This implies that $\rho(1450)$ is only weakly
affected by spontaneous breaking of chiral symmetry, in contrast
to the lowest baryon states. This indicates that the physics of
the  excited meson states and of the lowest baryons is very
different and that perhaps we observe a transition to the restoration
of chiral symmetry in highly excited hadrons \cite{Gloz1,Gloz2}.

These are important conclusions and we expect that the application of our
source techniques in a large scale study can lead to a deeper understanding 
of these issues from a lattice perspective. 

\begin{acknowledgements}
We want to thank Dirk Br\"ommel and Meinulf G\"ockeler
for valuable discussions, and Stefan Schaefer for important contributions
in the earlier stages of the BGR collaboration. The calculations were done 
on the Hitachi SR8000 at the Leibniz Rechenzentrum in Munich and we thank 
the LRZ staff for training and support. We acknowledge support by 
Fonds zur F\"orderung der Wissenschaftlichen Forschung in \"Osterreich,
projects  P16310-N08 and P16823-N08   
and by DFG and BMBF.
\end{acknowledgements}

\end{document}